\documentstyle[12pt]{article}
\begin{document}
\begin{center}
{\Large \bf Positivity constraints in QCD}

\bigskip

{\large O.V. Teryaev}
\date{}

\smallskip

{\it Bogoliubov Laboratory of Theoretical Physics,
Joint Institute for Nuclear Research,
141980 Dubna, Moscow region,  Russia}

\end{center}

\begin{abstract}
The density matrix positivity is a natural counterpart of unitarity.
The resulting constraints for various parton distribution
and correlations are considered.
Their compatibility with leading order QCD evolution is guaranteed
by the probabilistic interpretation of the latter. In the case of
non-forward distributions the positivity constraints naturally
imply the symmetric form of the evolution equation.
\end{abstract}



The positivity of density matrix is just the positivity of its
eigenvalues
\begin{eqnarray}
\lambda_i \geq 0,
\label{p}\end{eqnarray}
required, together with the normalization condition
\begin{eqnarray}
\sum _i \lambda_i =1,
\label{n}\end{eqnarray}
by its standard probabilistic interpretation. As inequalities (\ref{p})
may be written in terms of quantities $Tr \rho^i$ \cite{min}
\begin{eqnarray}
Tr \rho^2 \leq 1, \nonumber \\
2 Tr \rho^3 -3 Tr \rho^2 \geq  1, \nonumber \\
.....................
\label{pos}\end{eqnarray}
their preservation during interaction is guaranteed by the
$S-$ matrix unitarity, like the probability conservation (\ref{n}).
Positivity is a natural counterpart of unitarity. It plays the more
essential role, the more eigenvalues of density matrix are important.

The positivity in QCD is pronounced in the framework of factorization,
containing the nonperturbative ingredient -- parton distributions.
They may be considered (at leading order) as a density matrices
of partons in hadrons. The simplest application of positivity is
resulting in the well known constraint for spin-dependent and
spin-averaged distributions:
\begin{eqnarray}
f_+(x), f_-(x) \geq 0,
\label{+}\end{eqnarray}
or
\begin{eqnarray}
|\Delta f(x)(=f_+(x)-f_-(x))| \leq f(x)(=f_+(x)+f_-(x)) \geq 0.
\label{pos2}\end{eqnarray}

The non-diagonal (in helicity) elements of density matrix are also
constrained, the well-known example being provided by the case of
Soffer inequality \cite{S} for quark transversity
distribution:

\begin{eqnarray}
|h_1(x)| \leq q_+(x) = {1 \over 2} [q(x) + \Delta q(x)].
\label{S}\end{eqnarray}

The similar inequality \cite{longl} in the case of gluons
is relating the contribution of twist 2 ($G(x)$), 3 ($\Delta G_T(x)$)
and 4($G_L(x)$):
\begin{eqnarray}
\label{ineq}
|\Delta G_T(x)| \leq \sqrt{1/2G(x)G_L(x)}~.
\end{eqnarray}
It is most instructive to use this relation to estimate
$G_L$ (being the interesting new ingredient of nucleon structure
\cite{GI}) from below:
\begin{eqnarray}
\label{bound}
G_L(x) \geq 2[\Delta G_T(x)]^2/G(x).
\end{eqnarray}
This bound is analogous to
well-known condition established long time ago by
Doncel and de Rafael \cite{DDR}, written in the form
\begin{eqnarray}
\label{DDR}
|A_2| \leq \sqrt{R}~,
\end{eqnarray}
where  $A_2$ is the usual transverse asymmetry and
$R=\sigma_L/\sigma_T$ is the standard ratio in DIS.

The similar bound relating the contributions of different twists may be
derived for the twist-3 single spin asymmetries \cite{ET84}.
Taking into account
the so-called gluonic poles \cite{SQ} is resulting in the twist-3 term
which is behaving like $1/(1-x)$ with respect to the twist-2
unpolarized cross-section. This would necessarily mean the existence
of twist-4 correction, behaving, in turn, like $1/(1-x)^2$.
As a result, the $x-$dependence of the asymmetry is governed by the
generic equation \footnote{In the case of single asymmetry of hadronic
collisions, when the integration over momentum fraction of the unpolarized
hadron is present, the average values of the coefficients $a$ are
assumed. If $x_F \to 1$ they are close to the values at $x=1$}.
\begin{eqnarray}
\label{SSA}
A ={{a_3/(1-x)}\over{a_2+a_4/(1-x)^2}}={{a_3}\over{(1-x) a_2+a_4/(1-x)}}.
\end{eqnarray}
One can easily deduce two qualitative features of this result.

i) The maximal asymmetry is achieved when the contributions to the
unpolarized cross section of twists 2 and 4 are equal to each other,
indicating the possible large contributions of higher twists.

ii) This asymmetry (keeping twist 2 and 4 terms in the denominator)
is twice smaller with respect to its "naive"
estimate, when just the ratio of twist-3 and twist-2 terms is calculated.

The same conclusions are valid, when $P_T$ dependence of the asymmetry is
considered,
\begin{eqnarray}
\label{SSAP}
A ={{b_3 M/p_T}\over{b_2+b_4 M^2 /p_T^2}}=
{{b_3 }\over{b_2 p_T/M +b_4 M /p_T}}
={{b_3 M p_T }\over{p_T^2 b_2+M^2 b_4}}.
\end{eqnarray}
where we recovered also the standard form of $p_T$ dependence.
While maximal asymmetry is still estimated, up to a factor 2,
by the "naive" asymmetry, its slope at small $P_T$
governed by the twist-4 term. This is applicable for both
fermionic and gluonic poles cases. Note that simultaneous change
of $x$ and $p_T$ may be considered resulting in the same qualitative
conclusions i) and ii).
One may add to the positivity properties of the observed asymmetries
the bounds for the matrix elements, which are most simple
for the gluonic pole case, when the matrix element $T(x,x)$
is relevant:
\begin{eqnarray}
\label{ineqt}
|T(x,x)| \leq \sqrt{q(x)T(x,x,0)},
\end{eqnarray}
where $T(x,x,0)$ is a twist 4 matrix element with 2 zero-momentum gluons.

The positivity bounds may be derived \cite{MR,PST,Ji2,Rad2}
also for the non-diagonal
in {\it momentum} elements of parton-hadron density matrix,
known as non-forward (off-forward) parton distributions \cite{Ji,Rad}.
They are rather similar to (\ref{S},\ref{ineq}), although their exact
form depends on the actual definition of such a distribution,
existing in several versions. We present here the inequality
for the quark off-forward distribution \cite{Ji}
\begin{equation}\label{q(geo)}
|H_q(x, \xi)| \le \sqrt
{{ q (x_1) q (x_2)} \over{1-\xi^2}}
\end{equation}
and gluon non-forward distributions \cite{Rad,MR}
\begin{equation}\label{g(geo)}
|x^{'} g(x_1,x_2)| \le \sqrt {x_1 x_2 g(x_1) g(x_2)}, ~~x^{'}=x-\zeta.
\end{equation}
Here $\xi=\zeta(2-\zeta)$ is a "skewedness" parameter,
while variables $x_{1,2}$
measure the parton momentum fraction with respect to {\it different}
hadron momenta. Their relation to the standard notation are
different for off-forward ($x_{1,2}=(x \pm \xi)/(1 \pm \xi)$)
and non-forward ($x_1=x, x_2=(x-\zeta)/(1-\zeta)$) distributions.
These variables, as was noted in \cite{PST}, are suitable for expressing
the symmetry \cite{MPW} resulting from T-invariance (and similar to
the emerging in the case of forward twist-3 correlations \cite{ET84}),
\begin{eqnarray}
x'g(x,x') \equiv M (x_1,x_2)=M (x_2,x_1),
\label{m}\end{eqnarray}
(where $M(x_1,x_2)$ is the non-forward momentum distribution)
and for $t=ln Q^2$ evolution:
\begin{eqnarray}
{d M(x_1,x_2) \over{dt}}={\alpha_s \over {2 \pi}}
[\int_{x_1}^1 {dz\over{z(1-z)}} \tilde P(z,z^{'})
M (x_1/z,x_2/z') - \nonumber \\
{{M (x_1,x_2)}\over 2}
(\int_0^1 {dz\over{1-z}} \tilde P(z)+\int_0^1 {dz'\over{1-z'}}
\tilde P(z^{'}))].
\label{evm}\end{eqnarray}
Here symmetric evolution kernel is related to the standard one \cite{MR}:
$\tilde P(z,z^{'})=\tilde P(z',z)=z'(1-z) P(z,z^{'})$,
implying also the similar
relation for diagonal kernel: $\tilde P(z,z) = \tilde P(z)=z(1-z) P(z)$.
Only the regular parts of the kernels are present here, while the
singular contribution is written down explicitly.
To verify the symmetry of the
evolution equation one should note that new variables lead to the simple
relation between the integrations over $z$ and $z'$
\begin{eqnarray}
\int_{x_1}^1 {dz \over{z(1-z)}}...=\int_{x_2}^1 {dz' \over{z'(1-z')}}...,
\label{jac}\end{eqnarray}
implied by the $t$-channel momentum conservation
\begin{eqnarray}
{{x_1} \over{(1-x_1)z}} - {{x_2} \over{(1-x_2)z'}}=
{1 \over{1-x_1}}-{1 \over{1-x_2}}.
\label{int2}\end{eqnarray}
Such a role of variables $x_1,x_2$ is not occasional, as they
make non-forward distribution look like $\sum AB^*+A^*B$ (while forward ones
are $\sum A^2, \sum B^2$) \cite{PST}
and naturally manifest hermiticity, being the essential
ingredient of the symmetry properties. The evolution equation,
in turn, should present this symmetry.
It was stressed \cite{Rad2} that in the case of the double
distribution this symmetry is manifested,
provided the overall factor $1-\zeta/2$ \cite{MPW} is extracted.
However, the symmetry of nonforward distribution is more complicated
in that case.
One can see that
\begin{eqnarray}
F_\zeta(X)={1 \over{1-\zeta}} F_{-{{\zeta}\over {1-\zeta}}}
({{X-\zeta} \over{1-\zeta}}).
\label{sf}\end{eqnarray}
At the same time, the standard non-symmetric choice \cite{MR}
is leading to the cancellation of the overall factor $1/(1-\zeta)$ in the
r.h.s., so that resulting symmetry is just the interchange of $x_1$
and $x_2$, as was shown above.

The obtained evolution equation allows one to prove the stability
of the positivity constraint (\ref{g(geo)}) against $Q^2$-evolution,
following the general line of \cite{BLT,BST}. To do so, one
should consider the positive quantities (at some initial scale $Q_0$)
$M_{\pm}(x_1,x_2)=a M (x_1) + M (x_2 )/a \pm 2 M (x_1,x_2)$,
where $a$ is an arbitrary positive number.
The inequality (\ref{g(geo)})
\begin{equation}\label{g(geo1)}
|M (x_1,x_2)| \le \sqrt {M (x_1) M (x_2)},
\end{equation}
is just the result of minimization with
respect to its variation, where $M(x)=xg(x)$
is a diagonal momentum distribution.
Making use of (\ref{jac}), one may write
the evolution equations like:
\begin{eqnarray}
{d M_{\pm}(x_1,x_2) \over{dt}}={{\alpha_s} \over {2 \pi}}
[\int_{x_1}^1 {dz \over{z(1-z)}} (a \tilde P(z) M(x_1/z)+\tilde
P(z') M(x_2/z')/a \nonumber \\
\pm 2 \tilde P(z,z^{'}) M (x_1/z,x_2/z')) - M_{\pm} (x_1,x_2)
\int_0^1 {dz\over{1-z}} \tilde P(z)].
\label{evm+}\end{eqnarray}
It is very important that the virtual contributions are {\it diagonal}
in index $\pm$, so that they cannot change the positivity of the
distribution (when the distribution gets too close to zero,
it stops decreasing) like the exponentially decreasing positive function,
which cannot change sign \cite{BLT,BST}\footnote{Such property of
virtual correction is making
the positivity preservation especially simple when
the evolution in $x$ space is considered, while in the $N$ space it
requires the more elaborate analysis \cite{Nacht}, and the preservation
of, say, Soffer inequality is conspired \cite{Jaf}.}.
To prove positivity
of the real term (c.f. \cite{BLT,BST})
it is sufficient to consider the minimization with respect to the
variation of $z$-dependent (positive) $a$,
which can only make the sum of two positive
diagonal terms smaller, than in the actual case of minimization with respect
to constant $a$:
\begin{eqnarray}
&\min_a \int_{x_1}^1 {dz \over{z(1-z)}}
(a \tilde P(z) M(x_1/z)+\tilde P(z') M(x_2/z')/a
\pm 2 \tilde P(z,z^{'}) M (x_1/z,x_2/z')) \geq \nonumber \\
&\int_{x_1}^1 {dz\over{z(1-z)}} \min_{a(z)} (a(z) \tilde P(z) M(x_1/z)+
{{\tilde P(z') M(x_2/z')} \over{a(z)}}
\pm 2 \tilde P(z,z^{'}) M (x_1/z,x_2/z'))= \nonumber \\
&2 \int_{x_1}^1 {dz\over{z(1-z)}} (\sqrt {\tilde P(z) M(x_1/z)
\tilde P(z') M(x_2/z')}
\pm \tilde P(z,z^{'}) M (x_1/z,x_2/z')).
\label{az}\end{eqnarray}
Writing down (\ref{g(geo1)}) for $x_1 \to x_1/z, x_2 \to x_2/z'$,
the sufficient condition of positivity of (\ref{az}) can be
easily found \cite{PST}:
\begin{equation}\label{g(p)}
|\tilde P (z,z')| \le \sqrt{\tilde P (z) \tilde P (z')}.
\end{equation}
Such inequality is really valid \cite{PST} for all $z,z'$,
completing the proof of positivity. Although we considered here a
pure gluodynamics, the mixing is improving the situation
with positivity, providing extra positive terms, like in a forward case
\cite{BLT}.

One may consider the positivity of $Q^2$-evolution in an
"opposite" manner, when the general positivity constraints
for various parton distributions may be used for obtaining bounds
for yet unknown evolution kernels. From this point of view,
the bounds (\ref{bound},\ref{DDR}) are most interesting, putting
the constraints for twist-4 evolution. It is interesting, that
the evolution of $g_T(x)=g_1(x)+g_2(x)$, proportional to $A_2$,
combines the twist-2 (from $g_1$ and Wandzura-Wilczek part of $g_2$)
and twist-3 terms, the latter, in turn, being multiplicative
only in the limits $x \to 1$ and $N_C \to \infty$ \cite{ABH}.
However, the collinear singularities of {\it real} contribution to $g_T$
(essential for positivity properties) after making use of symmetry
due to T-invariance, are resulting in multiplicative renormalization.
The virtual corrections should be cancelled in the non-diagonal
in $\pm$ terms in evolution of $g_T(x) \pm a f(x)+f_L(x)/a$ (c.f.
\ref{evm+})).

The gluonic bound (\ref{bound}) may be used in a similar way
to estimate the twist-4 evolution.
Note that conclusion \cite{GI} about the absence of standard evolution
for longitudinal gluon distribution is, generally speaking,
applicable only for the effective twist-2 distribution, obtained
by the contraction with the box gluon-photon diagram,
the pole of which is cancelling the mass parameter, appearing, as usual,
in the twist 4 contribution.

The account for NLO effects is more peculiar. The preservation of
positivity for the parton distribution is depending on the choice
of the factorization scheme and may be considered as an extra constraint
for the latter \cite{BLT}. At the same time, full NLO result, defining
observable asymmetry, should respect positivity. Note, however, that
the attributing the positivity in QCD to the physical asymmetry \cite{Alt}
is, generally speaking, too strong condition. Say, Soffer inequality
is not related to any physical asymmetry. Instead, it may be obtained from
the positivity in the "gedanken-experiment" of the DIS scattering mediated
by the hypothetical currents \cite{Jaf}.
The one-loop coefficient functions of quark contributions (containing both
$q_+(x),h_1(x)$) to this process,
combined with the two-loop evolution kernels,
should guarantee the validity of Soffer inequality at NLO.
This would define the "gedanken-experiment" factorization scheme,
where the coefficient function of this process preserve its Born value
and Soffer inequality is valid for the distributions.
As to the "Drell-Yan" factorization scheme, it should definitely
preserve more weak inequality $|h(x)| \le q(x)$.

The inequalities (\ref{bound},\ref{DDR}) may be used also for estimation
of renormalon \cite{ren} contribution to twist-3 terms, the direct
application being impossible because the Born term
in the perturbative seriesis absent. Although positivity bound for $A_2$
in DIS is far from saturation, this is not necessarily the case for
the twist 3 terms in single spin asymmetries, and the obtained bound may be
of interest.

To conclude, the positivity constraints may be used for both
perturbative and nonperturbative ingredients of QCD factorization.
They should provide the bounds for parton distributions and
correlations, and for yet unknown evolution kernels.

\vspace*{7mm}
I am indebted to D.Boer, C. Bourrely, E. Leader, B. Pire,
J. Soffer and R. Tangerman for collaboration in the
studies of described problems, numerous discussions, assistance
and advice.

This research is partly performed in the framework of the Grant
96-02-17631 of Russian Foundation for Fundamental Research.

\end{document}